\documentclass[prb,aps,amsmath,amsfonts,amssymb,twocolumn, superscriptaddress,footinbib,showpacs, 10pt]{revtex4}
\usepackage{color}
\usepackage{amsmath}
\usepackage[pdftex]{graphicx}
\usepackage{subfigure}
\usepackage{epsfig}
\usepackage{soul}
\usepackage{ccaption}
\usepackage[justification=raggedright]{caption}
\newcommand{\nn}{\nonumber\\}
\newcommand{\beq}{\begin{equation}}
\newcommand{\beqn}{\begin{equation*}}
\newcommand{\eeqn}{\end{equation*}}
\newcommand{\eeq}{\end{equation}}
\newcommand{\barray}{\begin{eqnarray}}
\newcommand{\barrayn}{\begin{eqnarray*}}
\newcommand{\earray}{\end{eqnarray}}
\newcommand{\earrayn}{\end{eqnarray*}}
\newcommand{\balign}{\begin{align}}
\newcommand{\ealign}{\end{align}}
\newcommand{\ban}{\begin{align*}}
\newcommand{\ean}{\end{align*}}

\begin{document}

\title{The Influence Of Elastic Deformations On The Supersolid Transition}
\author{T.  Arpornthip}
\affiliation{Department of Physics, Washington University, MO 63160 USA}
\affiliation{Theoretical Division, Los Alamos National Laboratory, Los Alamos, New Mexico 87545 USA}
\affiliation{Department of Physics, University of Virginia, Charlottesville, VA 22904  USA}

\author{A.  V.  Balatsky}
\affiliation{Theoretical Division, Los Alamos National Laboratory, Los Alamos, New Mexico 87545 USA}
\affiliation{Center for Integrated Nanotechnologies, Los Alamos National Laboratory, New Mexico 87545 USA}
 \author{M. J. Graf}
\affiliation{Theoretical Division, Los Alamos National Laboratory, Los Alamos, New Mexico 87545 USA}
\author{Z.  Nussinov$^{1}$}

\date{\today}
\begin{abstract}
  We study, within the Ginzburg-Landau (GL) theory of phase transitions, how elastic deformations in a supersolid lead to local changes in the supersolid transition temperature. The GL theory is mapped onto a Schr\"odinger-type equation with an effective potential that depends
  on local dilatory strain. The effective potential is attractive for local contractions and repulsive
  for local expansion. Different types of elastic deformations are studied. We find that a contraction
  (expansion) of the medium that may be brought about by, e.g., applied stress leads to a higher (lower) transition temperature
  as compared to the unstrained medium.   In addition, we
  investigate edge dislocations and illustrate
  that the local transition temperature may be increased in the immediate vicinity of the dislocation core. 
  Our analysis is not limited to supersolidity. Similar strain effects should also play a role in superconductors.
\end{abstract}

\maketitle
\section{Introduction}

	Superfluids flow without resistance. 
	The existence of superfluidity raised the possibility of supersolids \cite{supersolid,
	Chester}- solids in which  superfluidity can occur without disrupting crystalline
	order. Long ago, 
	Chester \cite{Chester}
	theoretically demonstrated the possible existence of a supersolid. 
If supersolids exist, a natural contender would be solid Helium. \cite{Prokof'ev2007}
Recent torsional oscillator experiments \cite{KC} on solid
$^4$He pointed to supersolid  type features and have led to a flurry of activity. 
In the simplest explanation of the experiments, \cite{KC} a portion of the medium becomes,
at low temperatures, a superfluid that decouples from the measurement apparatus.
 Such a ``Non Classical Rotational Inertia'' (NCRI) effect
 is known to exist in superfluid liquid Helium \cite{London, Fairbank} 
 which was probed with similar techniques. \cite{Fairbank, Andron}
 Experimental results suggest the absence of superfluid features in
 ideal crystals with no grain boundaries.  \cite{Sasaki}
Currently, it is not clear if  an NCRI lies at the core of the recent experimental findings
 in solid $^4$He. For instance,
the required condensate
fraction adduced from a simple NCRI-only explanation
does not simply conform with thermodynamic measurements \cite{bgnt}. 
Rittner and Reppy \cite{Reppy} discovered that the putative 
super-solid type feature is acutely sensitive 
to the quench rate for solidifying the liquid. Aoki, Keiderling, and Kojima 
reported rich hysteresis and memory effects \cite{Kojima} similar to
those occurring in glasses \cite{memory}.  The torsional oscillator 
findings  can arise from material characteristics alone. \cite{dynamics, dorsey,Huse07,DSB,iw,andreev,korshunov,plastic} In particular, the thermodynamics and 
transient dynamics of distributed processes in amorphous or general 
non-equilibrated solids 
can currently fit \cite{bgnt,dynamics} observed results. 
Indeed, later  numerical results point towards
such a possibility. \cite{giuilio} Notably, recent experimental
results \cite{science} agree with an earlier suggested
theory concerning such transient dynamics. \cite{dynamics}
 The presence of non-uniformity in $^4$He is also suggested by a criterion 
 comparing the change in dissipation vs. relative period shift in torsion oscillator. \cite{Huse07}
 It may well be that these glassy and superfluid effects 
 are present in solid Helium. \cite{superglass}
 An interesting question concerns the coupling between elastic
 defects such as dislocations and superfluid type features. \cite{znmc}
 The coupling of the supersolid transition to impurities was 
 discussed in Ref.~\onlinecite{ba}.  The coupling between superfluidity and 
 elasticity in supersolids and how this may lead to a strain-dependent 
 critical temperature was discussed in Refs.~\onlinecite{DGT, Toner}. The viable existence of supersolid phase is
not confined to solid Helium. Other contenders for the supersolid state include cold atoms in a confining optical lattice. \cite{coldatom} There has been much work examining
supersolidity in spin systems as well, see, for example,  Ref.~\onlinecite{cristian}.
Supersolids constitute a fascinating
state of matter and appear in a host
of systems.

This article focuses on the coupling between nanoscale structure and 
supersolidity. \cite{DGT, Toner} As is well appreciated, elastic strain may fundamentally
affect local and mesoscopic electronic, magnetic
and structural properties. There is ample
evidence for significant coupling amongst the electronic
degrees of freedom with the lattice distortions in
cuprates, manganites, and ferroelectrics.  \cite{review-ela, strain_PRL}
The central thesis of this work is that elastic distortions may alter 
the supersolid behavior.   As we will elaborate later on, in, e.g.,  a cylindrical torsional 
oscillator geometry in which the boundary of the solid is elastically deformed so that
it undergoes a supersolid transition at a higher temperature than the bulk, 
a fraction of the boundary will become a supersolid  leading to
a partial decoupling of the bulk from the torsional oscillator chassis
and a consequent reduction in the period.

In this work, we will employ a Ginzburg-Landau
(GL) theory to study the influence of elastic 
strain on supersolidity. As we will show, the Euler-Lagrange equations
for the GL free energy result in an effective
Schr\"odinger type equation. We find the lattice distortion 
acts as an effective potential for the supersolid order parameter.
Solving the resulting effective Schr\"odinger--type equation, we find our
main results:  (1) a contraction (expansion) of the lattice edges leads
to an increase (decrease) in the local
supersolid transition temperature;
(2) elastic defects, such as dislocations,
lead to similar effects.

Although our motivation is the analysis of supersolids, 
all of our calculations within the GL framework
{\em are identical} for non-uniform elastically strained superconductors \cite{strain_PRL,sc1}
and lead to the same general conclusions, which we will derive
in this work. The case of uniformly strained superconductors 
has been investigated in detail in myriad experiments, starting
from Ref. ~\onlinecite{sizoo} and many works since. \cite{pressuresc}
It was found in these works that uniform hydrostatic pressure can increase the superconducting transition temperature.
The influence of pressure on the superconducting temperature has also been investigated in numerous theoretical treatments, e.g., Refs.~\onlinecite{ozaki, betouras}. 
Our GL formulation and Schr\"odinger- type equation give rise to an increase of the superconducting transition temperature
under applied pressure.

The outline of the paper is as follows:
in Section \ref{framework},
we set up the general GL framework for
our investigations. We illustrate the 
connection between the Euler-Lagrange equation
and the Schr\"odinger equation. In the sections
thereafter, we focus on particular lattice
distortion profiles to determine the change
in the local supersolid transition temperature.
In Section \ref{contract}, we examine the 
influence of a boundary edge contraction, and
in section \ref{expan}, we study the opposite
case of a boundary edge expansion.
In section \ref{edge_dis}, we analyze
the case of an edge dislocation. 
We summarize our findings in 
section \ref{conc}.

\section{General framework}
\label{framework}

	We study the GL free energy density
\beq\label{Fc}
	F(\vec{r})  = a(T)|\psi|^2  +  \frac{1}{2} b |\psi|^4 +c |\nabla\psi|^2 +\lambda(\vec{r})|\psi|^2,
\eeq
where $T$ is the temperature, $b$ and $c$ being positive constants,  $\psi$ the (complex)
supersolid order parameter, and $\lambda(\vec{r})$ a position dependent function
that captures the coupling of the order parameter to elastic strain as we elaborate on below. 
The prefactor $b$ in Eq.~(\ref{Fc})
is positive and depends only on the density of the crystal, as well as on defect densities. \cite{LLV9chp45}
For
temperatures $T<T_c$, the coefficient $a(T)$ is negative enabling a non-zero $\psi$ to
minimize the free energy. \cite{Tinkham1996chp4}  The condition $a(T_{c})=0$ determines the transition temperature $T=T_{c}$ below which supersolidity onsets. \cite{LLV9chp45}  
The third term 
in Eq.~(\ref {Fc}) relates the free energy with the magnitude of the gradient of $\psi$, as in a domain wall. \cite{Tinkham1996chp4}
	The difference between the free energy of a normal crystal and a displaced crystal appears in the last term.  
For a crystal whose constituents  $i$ undergo a distortion from an ideal unperturbed 
configuration $\vec{R}$ to a shifted configuration $\vec{R}^{\prime}$  due to 
the application of stresses, we set $\vec{u}_{i} = \vec{R}^{\prime}_{i} - \vec{R}_{i}$
and take the continuum limit wherein we replace $i$ by the continuous 
coordinate $\vec{r}$. In the up and coming, the Greek indices $\gamma,\delta$ 
will denote the spatial components
(e.g., $u_{\gamma = 1,2,3}$ will denote the Cartesian components of the displacement
$\vec{u}$ at site $\vec{r}$).
In general, a linear coupling of the form
	$a_{\gamma \delta} u_{\gamma \delta} |\psi|^{2}$ 
	is allowed between the linear order strain tensor $u_{\gamma \delta} =  \frac{1}{2} (\partial_{\gamma}
	u_{\delta} + \partial_{\delta} u_{\gamma})$  (where $\vec{u}$ is the elastic 
	displacement) \cite{kleinert}
	and the supersolid order
	parameter $\psi$. \cite{DGT, Aronovitz} 
	In what follows, we will consider, for simplicity, the case in which
	the displacement occurs only along one Cartesian direction.  Allowing
	for general displacements does not change our conclusions.
	For unidirectional displacements,  
	the coefficient of the last term in Eq.~(\ref{Fc}) can be expressed as
	a dilatory strain
\beq\label{lambda} 
\lambda(\vec{r}) = d  \vec{\nabla} \cdot \vec{u}(\vec{r}),
\eeq
where $d$ is a positive constant and $\vec{u}(r)$ is the displacement field. 
The sign of $d$ is chosen such that the free energy of Eq.~(\ref{Fc}) is lowered on 
introducing vacancies. The vacancy density scales with $[-(\vec{\nabla} \cdot \vec{u})]$
(whereas the interstitial density scales with $[(\vec{\nabla} \cdot \vec{u})]$).  Ions
in the vicinity of  a vacany will have an inward displacement towards
its location whereas ions in the vicinity of a interstitial will be pushed outwards. 
Eq.~(\ref{lambda}) and the free energy are functions of the strain tensor and thus
symmetric under spatial reflections under which $\vec{r} \to - \vec{r}$ 
and $\vec{u} \to - \vec{u}$. 
In bulk linear elasticity, the local strains scale, as in Hooke's law,  
as the pressure divided by the elastic moduli.
In the following sections, we will consider the strain fields associated with
various cases. 

As noted earlier, our GL theory of Eq. (\ref{Fc}) also describes a (singlet) superconductor
with an order parameter $\psi$ in the presence of elastic strains on which we comment below.  
In a charged crystal (of unit cell volume when undeformed), under applied elastic stress the electric field couples to the local charge density (which deviates 
from that of the undeformed crystal by an amount $[-(\vec{\nabla} \cdot \vec{u})]$ (and whose volume trivially scales as $(1+ \vec{\nabla} \cdot \vec{u})$)). 
For superconductors, $\lambda(\vec{r}) \to e^{*}  \phi(\vec{r})$ with $\phi$ the electrostatic
potential and $e^{*}$ an effective charge. \cite{znmc,diel}  
With Eq. (\ref{lambda}), the last  term in Eq.~(\ref{Fc}) is a general isotropic coupling between the strain and the supersolid (superconducting) order parameter. 
In the cases that we will examine the displacement  $\vec{u}$ will occur along one
Cartesian direction 
($\vec{u}$ will have only one  component). 
Furthermore, in the first two cases that we will detail below 
(contraction and expansion along an edge), this displacement field will vary only along one
Cartesian direction and will be uniform along all other orthogonal directions.
Consequently, the coupling $\lambda$ will depend only on one Cartesian
direction: $\lambda = \lambda(z)$.
In the last case discussed in this work, that of an edge dislocation,
the displacement field (and consequently the coupling $\lambda$)
will depend on two directions. 

	To find the ground state of such crystal, we want to minimize the free energy. The variational derivative of F with respect to \(\psi^{*}\) leads to the
Euler-Lagrange equation 
\begin{eqnarray}
\label{EL}
\frac{\delta F}{\delta \psi^{*}}	=	a(T) \psi + b |\psi|^2 \psi -c \nabla^2 \psi + \lambda(\vec{r}) \psi = 0,
\end{eqnarray}
with an identical (complex conjugated) equation for $\delta F/\delta \psi =0$.
In situations in which a weakly first- or second-order supersolid transition
occurs, we may, in the vicinity of the transition (where $\psi$ is
small) omit the cubic term in Eq.~(\ref{EL}),  \cite{quadrature} 
and the variational equation may be recast as
\begin{eqnarray}
	-c\nabla^2\psi + \lambda(\vec{r})\psi = -a(T)\psi.
	\label{cla}
\end{eqnarray}
Eq.~(\ref{cla}) is a Schr\"odinger type equation with $c = \hbar^2/2m$ and $a(T) = - E$
with $E$ the energy and $m$ a mass. 
Solving for the eigenvalue $E = -a(T)$ enables us
to extract the transition temperature. Generally, 
a shift in the transition temperature results
from the coupling to the elastic displacements.

The gradients of $\vec{u}$ as embodied in $\lambda(\vec{r})$, take on the role of a potential energy in the effective quantum problem for the ``wavefunction'' $\psi$.
We briefly comment that the case of {\em uniform pressure} corresponds to a constant $(\vec{\nabla} \cdot \vec{u})$ and thus to a constant effective
potential $\lambda(\vec{r}) = const$. Applied to the analysis to be presented below to more complicated cases, 
such a uniform shift of the potential energy (and thus to the eigenvalues $E$) leads to a constant shift in the value of $a(T)$ at the transition point. 
As $a(T)$ is monotonic in temperature, for $d>0$ ($d<0$) this leads to an increase (decrease) in the transition temperature for a uniform contraction ($\vec{\nabla} \cdot \vec{u} <0$) as it may indeed occur under uniform applied pressure in superconductors for which for an increase or decrease of the superconducting $T_c$ appear for
different systems. \cite{sizoo,pressuresc} 
The case of a spatially uniform dilatory stress is a particular simple limiting form 
of the more general non-uniform elastic deformations that we discuss in this work. 
	
	In the remainder of this work, we will examine the
	solutions of Eq.~(\ref{cla}) for various non-uniform elastic displacements $\vec{u}$. 
	In particular, we will examine the strain fields associated with
	a contraction of the sample boundaries, an expansion of a boundary
	edge, and the strain profile associated with an edge dislocation.
	
\section{Contraction of boundary edge}	
\label{contract}

\begin{figure*}[bt]
				\includegraphics[width=0.6\textwidth]{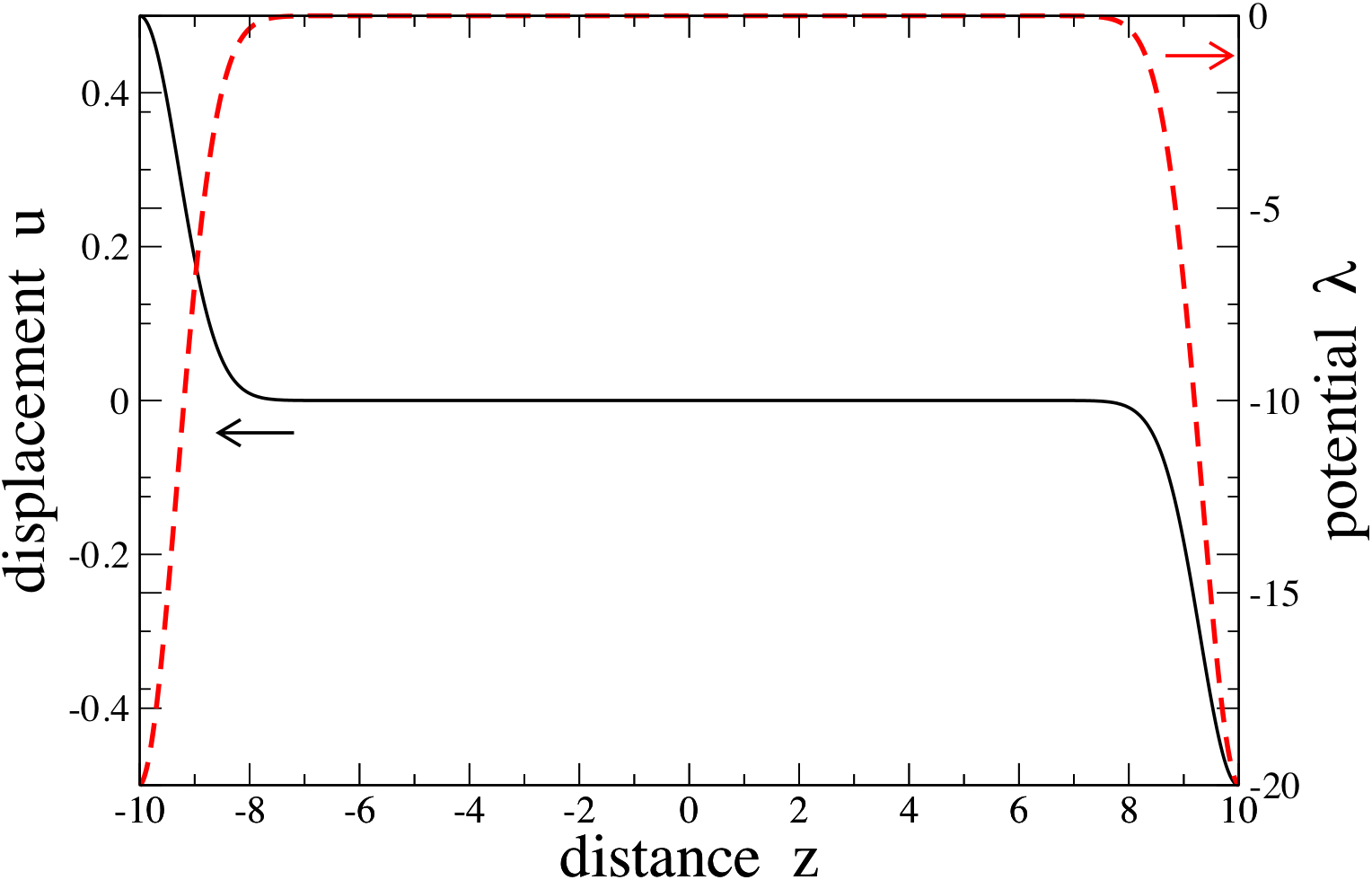}
	\caption{(Color online.)
 The displacement corresponding to a contraction near the edges. For clearly illustrating the contraction,
 we depict exaggerated displacements. In this sketch, the displacement (black solid) is given by Eq.~(\ref{Uc}) with $L=20, u_0=0.5, d=1$, and $k=1$ where the lattice constant set to unity. 
 The effective potential (red dashed) is given by Eq. (\ref{Wc}).
 The large (exaggerated) value of the displacement  $u_{o}$ is chosen to vividly illustrate the contraction.}
 	\label{fig:ContractedCrystal}
\end{figure*}

	Consider a crystal with a side of length $L$ along one of the Cartesian directions
	(the coordinate values corresponding to this side are 
	in the range $L/2 \ge z \ge -L/2$). We consider a contraction in which near the two edges, 
	the lattice sites are most displaced from their equilibrium positions, see  Fig.~\ref{fig:ContractedCrystal}.
	Such a contraction may, e.g., be brought about by applying stress (along opposite directions) on the two edges
	of the system. 
	Alternatively, a shock wave or generation of coherent phonon propagation  by  ultrafast pump-probe spectroscopy may be used to create local density modulations resulting in nonuniform strain near the edges of the sample.\cite{UOS}
As we will show, the displacement at the edges leads to a change in the local transition temperature.
A displacement field describing a contraction along the $z$ direction is given by 
\beq\label{Uc}
	u_{z} = \begin{cases}u_0[e^{-(z+L/2)^2/k^2}-e^{-(z-L/2)^2/k^2}] & \mbox{ for } |z| \leq L/2 \\ 0  & \mbox{ for } |z| > L/2 \end{cases}.
\eeq
The displacement thus occurs in some finite region (of scale $k$) about the edges.
$u_0$ the maximum displacement, and with no
displacement along the $x$ or $y$ directions, $u_{x}= u_{y} =0$.

The corresponding effective potential of Eq.~(\ref{lambda}) is given by
\beq\label{Wc}
\lambda =  \frac{2 u_{0} d}{k^{2}} \Big[ (z- \frac{L}{2}) e^{-(dz-\frac{L}{2})^{2}/k^{2}} - (z+ \frac{L}{2}) e^{-(z+ \frac{L}{2})^{2}/k^{2}}\Big].
\eeq
For $|z|>L/2$ (points outside the crystal), 
the supersolid order parameter $\psi  = 0$ and in Eq. (\ref{cla}) the effective potential  $\lambda = \infty$.  
For small deformations, this attractive potential leads to the appearance of a weak bound state.
For $z>0, ~ L/2 \gg (L/2-z) \gg k \sqrt{\ln (2u_{0}d/k)} \equiv \epsilon/2$, the effective potential tends to zero,
and the bound state wavefunction is of the form $\psi \sim \exp[\kappa (z-L/2)]$. A similar form
is attained near the point $z= -L/2$. The value of
$\kappa$ and thus of the bound state energy $E = - c \kappa^{2}$ can be computed 
in the standard way by integrating 
the Schr\"odinger equation once in a region of width $\epsilon$ across the point $z=L/2$ in an
extension
of the problem to $z>L/2$ in which the potential is symmetrized about the point $z=L/2$. As
$|E| \ll |\lambda(z)|$ in the narrow region near the edges, Eq.~(\ref{cla}) reads
\beq\label{intSE}
	 - 2 \kappa = \left[\frac{d\psi}{dz}\right]^{L/2+\epsilon/2}_{L/2-\epsilon/2} =
	\frac{1}{c} \displaystyle\int _{L/2 -\epsilon}^{L/2 + \epsilon} \lambda(z) dz.
\eeq
Ignoring exponentially small corrections, we attain that the bound state energy is
\begin{eqnarray}
E = -\frac{d^2u_0^2}{4c}.
\label{Ee1}
\end{eqnarray}
We will now employ the {\em value of E to determine  a change in the transition temperature}. Within the 
	GL theory,  $a(T)  \simeq \alpha(T-T_c^0)$ near the transition temperature, where $T_c^0$ is the unaltered transition temperature and $\alpha>0$ is a constant.  Writing $a+E = \alpha(T-T_{c}^{eff})$
	where $T_c^{eff}$ is the effective transition temperature, we have
\beq\label{Tc}
T_c^{eff}= T_c^0 + \Delta T_{c},
\eeq
with 
\begin{eqnarray}
\label{Tc1}
\Delta T_{c} =  \frac{d^2u^2_0}{4\alpha c}.
\end{eqnarray}
	In other words, the region near the contracted edges has {\em a higher transition temperature into the supersolid
	state than the bulk.}  
	Generally, the maximal displacement
in Eqs.~(\ref{Uc}) and (\ref{Ue})  can be of order $u_{0} \sim 0.1$ lattice constants as set by the Lindemann criterion of melting
in most materials (or of $u_{0} \sim 0.2$ in solid $^{4}$He and potentially
other quantum solids). \cite{lind} In Eqs.~(\ref{Fc}) and (\ref{lambda}), the 
parameters $c, d = {\cal{O}}(1)$.  We estimate a small enhancement of the transition temperature in the surface region. For parameters $\alpha = 1/T_{c}^{0}, d = 1, u_0 = 0.1$, and $ c = 1$, we find from
Eq.~(\ref{Tc1}) a small enhancement compared to the bulk transition,
$\Delta T_c  = 2.5 \times 10^{-3} T_{c}^{0}$.

The effect of this shifted transition temperature is that, when a sample of contracted $^4$He is cooled down, the region near the edges would turn into supersolid at a higher temperature than the bulk of the crystal.  Perusing the form of the supersolid order parameter $\psi$,
	and Eq.~(\ref{intSE}),
	we see that $\psi$ drops exponentially away with the boundary
	with a penetration depth $\ell = 2c/(du_{0})$. 
With our previous estimates for parameters in Eq. (\ref{Tc1}) combined, 
we find that the penetration depth $\ell \sim 20$ lattice constants.

\begin{figure}[hbt]
	\includegraphics[width=0.3\textwidth,height=0.36\textwidth]{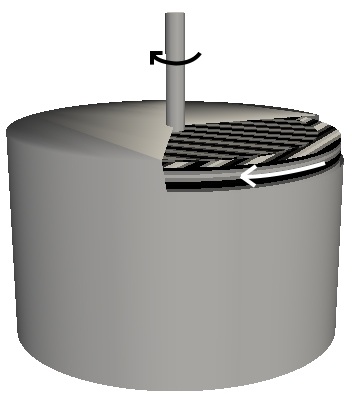}
	\caption{Under compression
	of the edges of the torsional oscillator,  the rim attains a supersolid component at a higher temperature
	than the bulk does. On cooling down to this temperature, the supersolid fraction of the
	rim partially decouples from the bulk and outer chassis.}
	\label{fig:decoupling-side}
\end{figure}

	 Returning to the NCRI \cite{KC, London, Fairbank, Andron} briefly discussed in the introduction, if the entire sample is rotating before the transition to a supersolid phase occurs, at some temperature higher than the normal transition to supersolid of bulk helium but low enough to make the edges become supersolid, the supersolid component in the edges will partially decouple from the bulk rotation. This situation is depicted schematically in Fig.~(\ref{fig:decoupling-side}).

	We now expand on the relation between the local value of the supersolid order parameter 
	$\psi$ and the local effective transition temperature. 
	Eq.~(\ref{cla}) holds for all locations $-L/2 \le z\le L/2$ (and trivially, of course,
	on any local segment within this region). We earlier solved Eq.~(\ref{cla}) to find that the supersolid order parameter decays exponentially with a decay distance $\ell$ away from 
	the boundary points $z = \pm L/2$.  Thus, deep within the bulk, the supersolid order
	parameter was zero. We may examine Eq. (\ref{cla}) locally (with
	a local effective potential $\lambda(z)$) in order
	to see when we may attain a finite supersolid order parameter $\psi(z)$ at general
	locations $z$ away from the boundary.  	
	
\begin{figure*}[tbh]
	\subfigure[]{\centering\label{fig:Teff_Crystal}
	\includegraphics[width=0.46\textwidth]{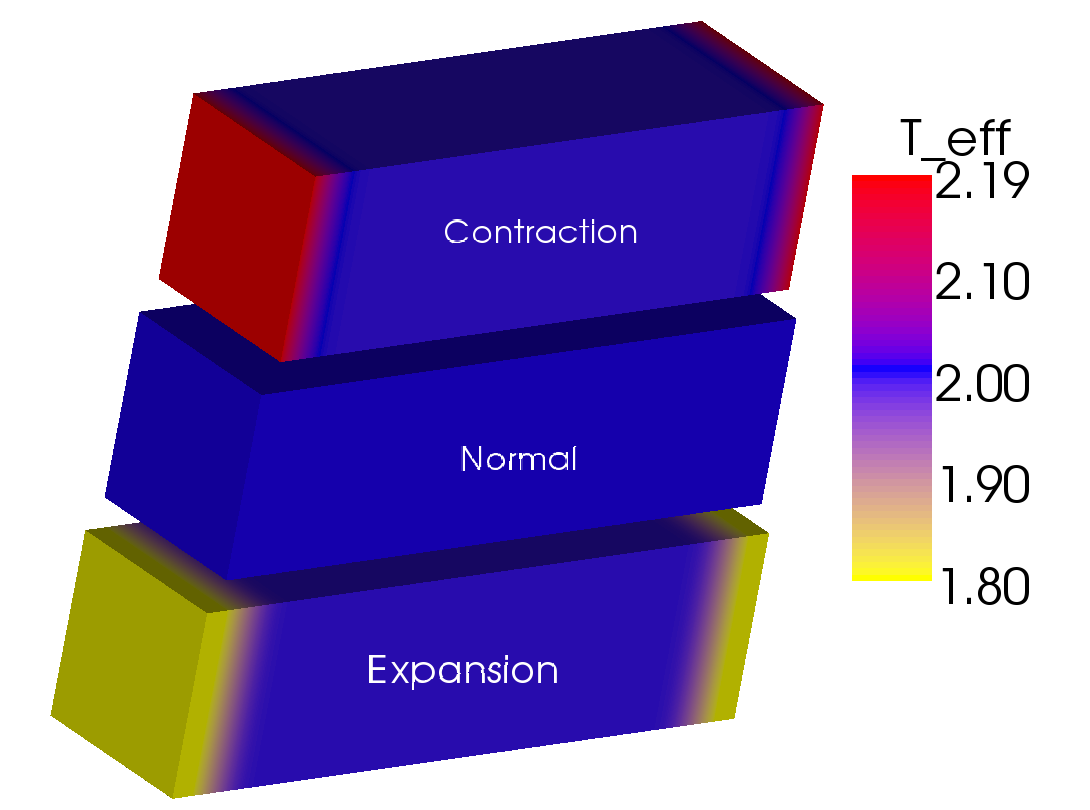}
	}
	\hfill
	\subfigure[]{\centering\label{fig:Teff_Plot}
	\includegraphics[width=0.46\textwidth]{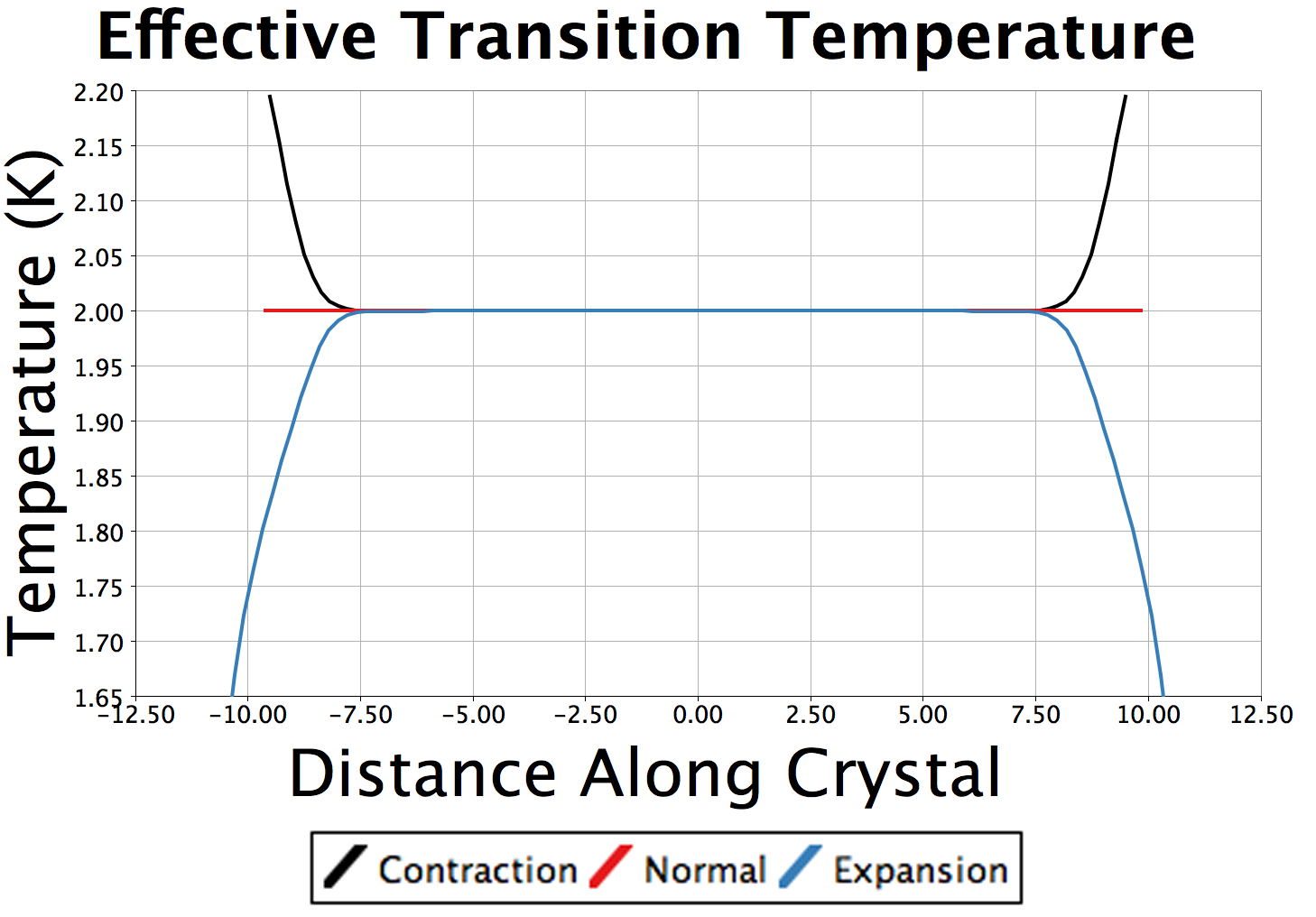}
	}
	\caption{(Color online.)
	 (a) The transition temperature along a slab is plotted in a contour map for different cases from top to bottom:  a slab with contraction, a uniform slab, and a slab with expansion near its edges. Whenever the elastic deformations are present, the local supersolid transition temperature is altered by comparison to the uniform solid. Near the edges, where the 
	elastic deformation is present, the supersolid transition temperature is altered: $T_{c}$ increases 
	at the boundaries in the case of boundary contraction and decreases for an expansion near the boundaries. The dark (black) line is associated with contraction, red line with the normal crystal, and blue with the expansion. With the lattice constant to unity, the parameters 
$k=1, T_c^0 = 2$ K, $u_0=0.5$ and $\frac{d^2 u_0^2}{4\alpha c} = 0.2$ K. 
(b) We plot the effective transition temperature for an undeformed crystal (straight line),
and that with a compression/expansion of its boundaries.} 
	\label{fig:Teff}
\end{figure*}

	As the displacement only occurs near the edges, and since the change in transition temperature is the result of the displacement, it is reasonable to assume that the change in transition temperature can only be detected in the region near the edges. For the region inside the crystal (far from the edge), the transition temperature should remain unaltered. Based on the observation, the transition temperature as a function of the z-axis (the axis parallel to the length of the crystal) could be described as
\beq
\label{tcc}
T_c^{eff}(z)  	=  T_c^0 + f(z) \left(\frac{d^2u^2_o}{4\alpha c} \right),
\eeq
where $f(z)$ is a function that rapidly varies from $1$ at the boundaries $z = \pm L/2$  to zero
for positions removed from the boundaries. An example is provided by
\beq\label{fz_ex}
f(z) = e^{-(z-L/2)^2/k^2}+e^{-(z+L/2)^2/k^2}.
\eeq
A contour plot of 
$T_c^{eff}$ is depicted in figure \ref{fig:Teff},
for $L=20$ (we set the lattice constant to be unity), 
$k=1, T_c^0 = 2$ K, $u_0=0.5$ and $\frac{d^2 u_0^2}{4\alpha c} = 0.2$ K. 
The large (exaggerated) value of the displacement  $u_{o}$ is chosen to lucidly 
illustrate a contraction as in, e.g., Fig.~\ref{fig:ContractedCrystal} and its effect.

\section{Expansion of edge boundaries}
\label{expan}

\begin{figure*}[bt]		\includegraphics[width=0.6\textwidth]{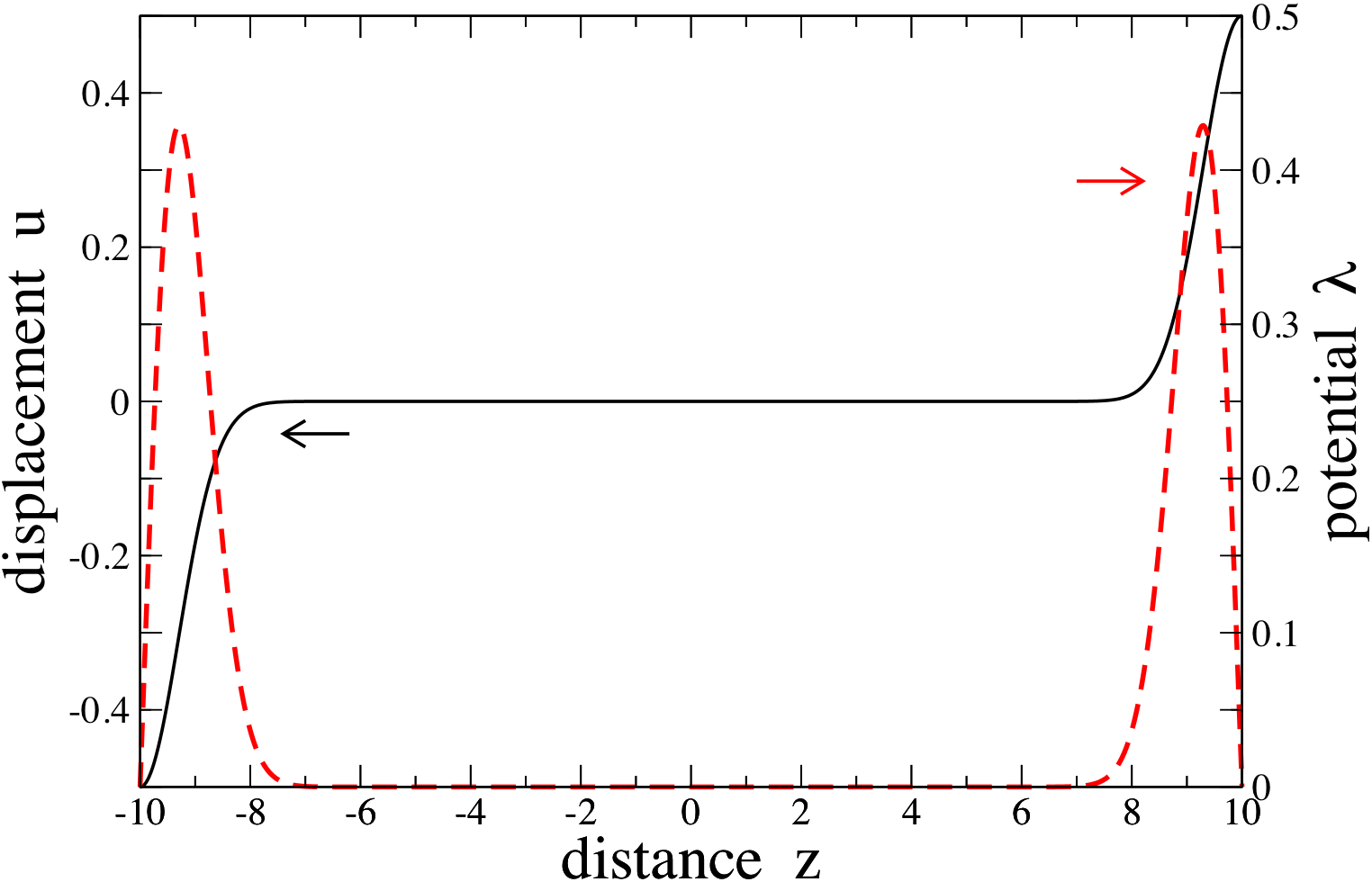}
	\caption{(Color online.)
	The displacement field corresponding to an expansion near the edges.  Plotted
	is the displacement field (black solid) given by Eq.~(\ref{Ue}) with $L=20, u_0=0.5,d=1$ and  $k=1$.
	The effective potential (red dashed) is given by Eq.~(\ref{Wc1}).
	The large (exaggerated) displacement highlights the expansion near the boundaries.}
	\label{fig:Expansion}
\end{figure*}

The situation of the expansion  near the edge boundaries is schematically shown in Fig.~\ref{fig:Expansion}.
As in the case of contraction, this may be physically brought about by applying opposite stresses  (e.g., shear stresses)
on the  two boundaries of the system. 
In an annular geometry similar to that in Fig. ~\ref{fig:decoupling-side}, an expansion may result
by a difference in pressures between the inner and outer parts of the cylinder. 
A typical displacement field $\vec{u}$ is, in this case, given by
\beq\label{Ue}
	u_{z} = \begin{cases}u_0[e^{-(z-L/2)^2/k^2}-e^{-(z+L/2)^2/k^2}] & \mbox{for } |z| \leq L/2 \\ 0  & \mbox{for } |z| > L/2 \end{cases},
\eeq
and $u_{x}= u_{y}=0$.
The variational equations give rise to a Schr\"odinger equation. The sign of $\lambda$ is flipped relative to the case of the contraction. In this case, $\lambda$ is everywhere positive reflecting a repulsive effective potential. This difference in sign gives rise to an important difference between expansion and contraction. In the case of expansion, the effective potential displays two peaks instead of two wells. In the absence of the two peaks, the problem reduces to that of a particle in an infinite potential well model. The wavefunction for the unperturbed ground state is now given by
\beq\label{PSIe}
\psi = \sqrt{\frac{2}{L}}\cos\left( \frac{\pi}{L}z \right).
\eeq
The energy of such a bound state in a box of size $L$ is
\beq\label{EUe}
E = \frac{\pi^2c}{L^2}.
\eeq

	Now, consider the perturbed state with the potential given by $\lambda = d \vec{\nabla} \cdot \vec{u}$ which reads
	\beq\label{Wc1}
\lambda =  \frac{2 d u_{0}}{k^{2}} \Big[ (z+ \frac{L}{2}) e^{-(z+ \frac{L}{2})^{2}/k^{2}} 
- (z- \frac{L}{2}) e^{-(z- \frac{L}{2})^{2}/k^{2}}\Big].
\eeq

	We may approximate $\lambda$ near its maxima  by delta functions.  The maxima occur at
	 $z = \mp\frac{L}{2} \pm \frac{k}{\sqrt{2}}$. We express $\lambda$ as
\beq\label{Lamdae}
\lambda = du_0\left[\delta[z-(\frac{L}{2}-\frac{k}{\sqrt{2}})] +\delta[z+(\frac{L}{2}-\frac{k}{\sqrt{2}})]\right].
\eeq
The first order approximation to the perturbed ground state energy trivially yields
\beq\label{Ee_temp}
E' = E + \displaystyle\int^{\infty}_{-\infty}\psi^*\lambda\psi dz =  E+ \frac{8du_0}{\pi}\sin\frac{k\pi}{L\sqrt{2}}.
\eeq
 Replicating the steps of Section \ref{contract},  we find that the effective transition temperature $T_c
	^{eff}$ for  the case
	of expansion is
	\beqn
T_c^{eff}(z)  	= T_c^0 - f(z)  \left(\frac{\pi^2c}{\alpha L^2} + \frac{8du_0}{\alpha \pi}\sin\frac{k\pi}{L\sqrt{2}}\right).
\eeqn
	In this case,  as the system is cooled down, the faces would become supersolid after the bulk crystal as we cool down the crystal. A plot is  given in Fig.~ \ref{fig:Teff} for
 $T_c^0=2$ K, $u_0=0.5, \frac{\pi^2c}{\alpha L^2} = 0.02$ K, and 
 $\frac{8d u_0}{\alpha\pi}\sin\frac{k\pi}{L\sqrt{2}}=1$ K. As in our prior analysis of
 the contraction, the large value of $u_{0}$ in Fig.~\ref{fig:Expansion}
 is chosen to vividly illustrate the elastic distortion associated
 with an expansion. 

It is worth highlighting the origin of the difference
between the cases of edge contraction and expansion.
Both cases have different divergences of the displacement
field (and thus different local density profiles). The local mass
or equivalently, the vacancy density is what couples to the supersolid order parameter. 
Note, in case of a superconductor it is the charge density that couples to the order parameter.
Both the displacement field and the spatial gradient
are odd under spatial reflection. In our case, $\vec{\nabla} \cdot \vec{u}$ 
is even under spatial reflection (it reflects
the scalar mass density) and the
two cases are physically very different even though
the spatial profile of the displacement fields in
both cases are related by a minus sign
(see Eqs.(\ref{Uc}) and (\ref{Ue})).

\section{Dislocations}
\label{edge_dis}
Below, we will 
present detailed numerical and variational calculations
of the local transition temperature due to an edge dislocation using
the formalism that we have employed thus far in this work. 
For a discussion of dislocations in the quantum arena see, e.g., Ref.~\onlinecite{znmc}.  
An analysis analogous to ours was done by Toner \cite{Toner} who reached similar
conclusions as we have. Some time after we discussed this phenomenon
\cite{us09}, Ref.~\onlinecite{filament} considered
the problem of dislocation line filaments which become supersolid
while the bulk is non-supersolid. This is markedly different from our approach
where both the bulk
and the dislocation core become supersolid at transition temperatures
that differ by small amounts. The small change in the ordering
temperature is imperative in our perturbative approach of linearly expanding $a(T)$
in Eq.~(\ref{Fc}) about the bulk supersolid transition temperature and 
in neglecting the cubic terms in Eq.~(\ref{EL}) when solving
the effective Schr\"odinger type equation of Eq.~(\ref{cla}).  

\begin{figure*}[tb]
	\subfigure[An edge dislocation]{\centering\label{fig:dislocation}
	\includegraphics[width=0.4\textwidth]{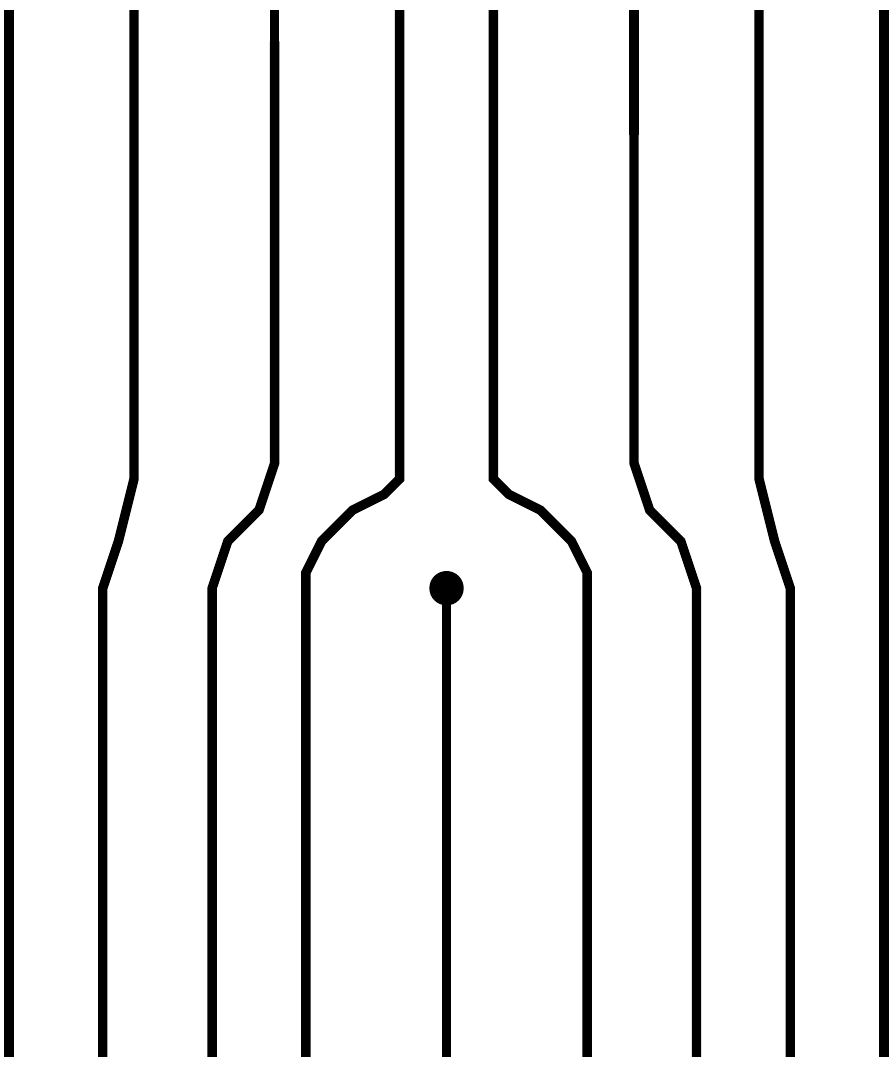}
	}
	\hfill
	\subfigure[The displacement field corresponding to an edge dislocation]{\centering\label{fig:vector}
	\includegraphics[width=0.45\textwidth]{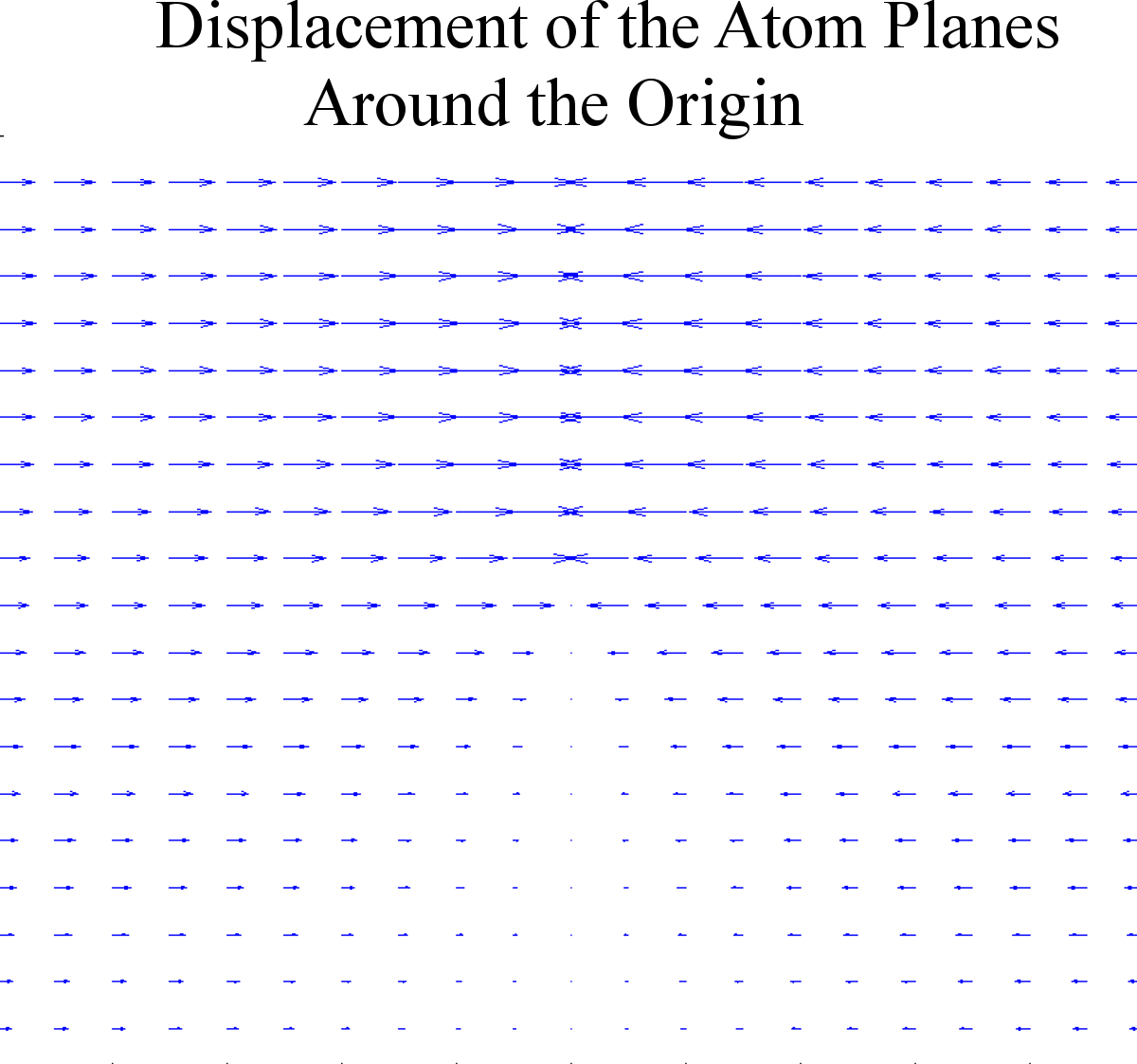}
	}
	\caption{A schematic of an edge dislocation. At left (a), shown are rows of atoms. The presence of an edge dislocation is manifest in the appearance of a different number of vertical rows of atoms above
	and below the terminal dislocation point. The corresponding displacement field is shown
	at right (b).}
\end{figure*}

Many displacement fields can correspond 
to a given ``Burgers vector''  $\vec{b}$ describing a dislocation.
The Burgers vector is defined by a circuit integral around a dislocation core 
\cite{kleinert} $b_{\gamma} = (\oint_{K} d \vec{s} \cdot \vec{\nabla} u_{\gamma})$ 
for a large contour $K$ around the dislocation core) describing a dislocation. 
We will analyze one such particular set of displacement fields.
All of these displacement fields are related to one another via a smooth
deformation $\vec{u} \to \vec{u} + \vec{v}$. Here, $\vec{v}$ is a non-singular 
vector field with a vanishing associated circulation:
 $\oint_{C} d \vec{s} \cdot \vec{\nabla} v_{\gamma} =0$
around any closed contour $C$. As we saw in the earlier sections, a smooth displacement 
field (corresponding to, e.g., a contraction or an expansion) can, on its own, 
raise or lower the effective local supersolid transition temperature. Thus,
the effective change in $T_{c}$, which we turn to next, will generally
depend on the detailed form of the displacement fields $\vec{u}$ corresponding
to a given dislocation. In what follows, we consider a particular minimal
displacement field form that corresponds to symmetric unidirectional displacements
about a lattice direction of an unstrained crystal.  With $x$ and $y$ denoting the 
horizontal and vertical Cartesian directions, we consider a particular displacement field in Figs.~\ref{fig:dislocation}) and \ref{fig:vector}
that corresponds to a dislocation with a Burgers vector $\vec{b} = b \hat{e}_{x}$.
In what follows, the spatial extent of the dislocation core will be set by $k$. 

\begin{figure}[th]
	\includegraphics[width=0.55\textwidth]{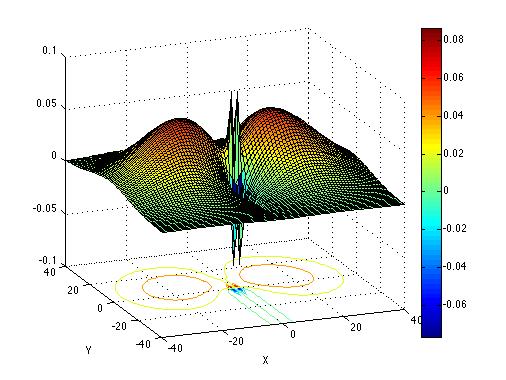}
	\caption{(Color online.)
	The effective potential energy of Eq.~(\ref{eq:potential})
	corresponding to the displacement function of 
	Eq.~(\ref{dislocation}). 
In this figure, $b=d=c=$ and $k=10$. That is,
$\beta =  0.0135$ and 
$\sigma = 0.0017$.
The maximum strain is 0.086 (in units of the lattice constant).}
	\label{fig:potential}
\end{figure}

	The following displacement
	field describes such an edge dislocation,
	\beq \label{dislocation}
	\vec{u}(x,y)  = - \frac{b}{2\pi}e^{-(x^2+y^2)/k^2}sgn(x) \cos^{-1} \left(\frac{-y}{\sqrt{x^2+y^2}}\right) \hat{e}_{x},
\eeq
where $sgn(x)$ is the sign of $x$, i.e., $sgn(x) = [2 \theta(x) -1]$
with $\theta(x)$ the Heavyside function. 
In Eq.~(\ref{dislocation}),
the magnitude of the Burgers vector $b$ cannot exceed 
the inter-atomic lattice spacing. Furthermore, realistically, $k$ 
may be of order of 10 (lattice constants).
We may derive an effective potential from the displacement in the same way we did for the above two cases
(Eq.~(\ref{lambda})). In this case, an analytical solution to Schr\"odinger equation is not possible and we will resort to a numerical solution. The effective potential energy is provided in Fig.~ \ref{fig:potential}.

\begin{figure*}[tb]
	{\centering\label{fig:initial}
	\includegraphics[width=0.45\textwidth]{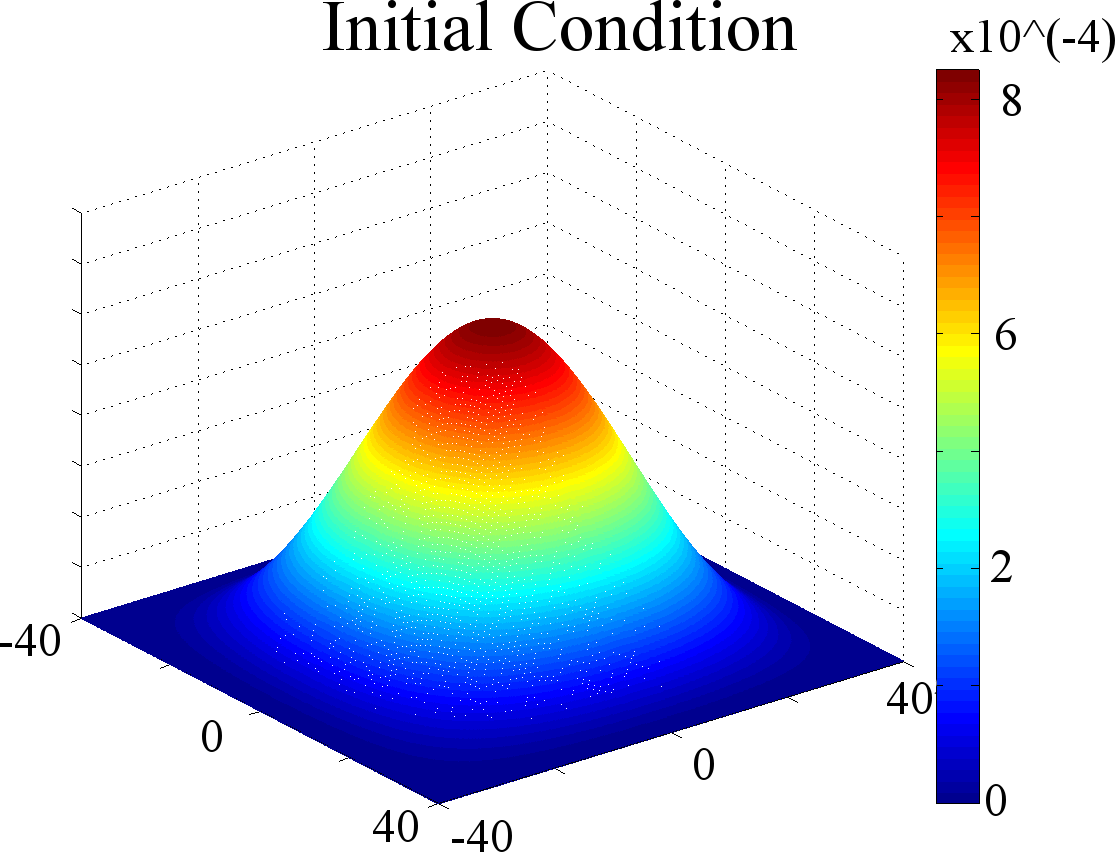}
	}
	\hfill
	{\centering\label{fig:ovr}
	\includegraphics[width=0.45\textwidth]{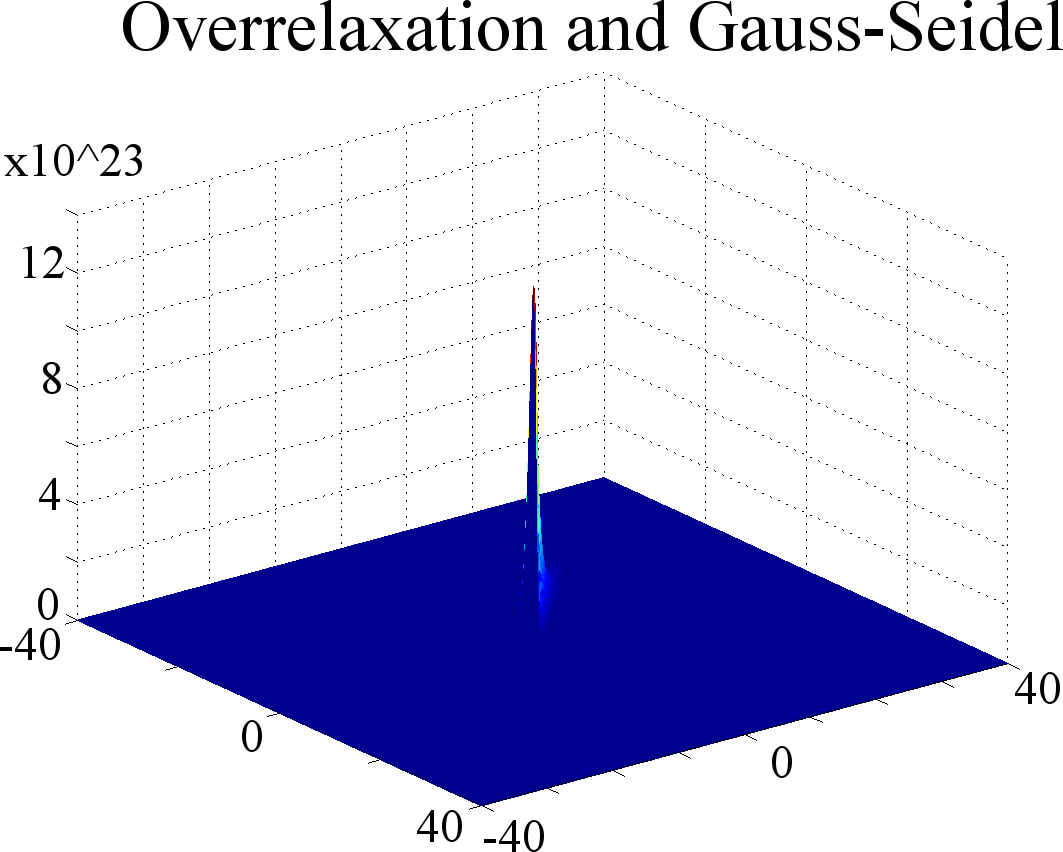}
	}
	\caption{(Color online.)
	The supersolid order parameter associated with the effective potential
	of Fig.~\ref{fig:potential}. Shown is the Gauss-Seidel solution of Eqs.~(\ref{EL}) and (\ref{cla}) with $c=1$ for the dislocation profile of Eq.~(\ref{dislocation}). Left:  An initial seed state. Right: The final ``wavefunction" (supersolid order parameter) $\psi$. The localized bound state of the order parameter is evident.}
	\label{fig:wavefunction}
\end{figure*}

We approximate the partial derivatives in the Schr\"odinger equation by finite differences and use the numerical Gauss-Seidel method for solving iteratively a system of linear equations in conjunction with over-relaxation.   Our relaxation scheme shows that the wavefunction localizes rapidly to the region near the origin where the dislocation core sits. Both the initial seed and the final numerical result are depicted in Fig.~\ref{fig:wavefunction}.
	The numerical solution to the Schr\"odinger equation illustrates that there is a change in transition temperature.

	We now turn to an approximate analytical solution. A contending variational state that is localized about the dislocation core 
is given by 
\begin{equation}
	\psi (x,y) = \sqrt{\frac{2 \sigma} {\pi}} e^{-\sigma(x^2+y^2)}	.
\end{equation}

From Eq. (\ref{dislocation}), we can compute the effective potential energy,
\begin{align} \label{eq:potential}
	V &\equiv \lambda  =  d (\vec{\nabla} \cdot \vec u) \nn
	  &= \frac{bd}{2\pi} sgn(x)e^{-\sigma(x^2+y^2)}\bigg[\frac{2x}{k^{2}} \cos^{-1} \left(\frac{-y}{\sqrt{x^2+y^2}}\right) \nn
	  &\qquad+ \frac {y}{x(x^2+y^2)}\bigg].
\end{align}
The Hamiltonian $ H = [-c \nabla^2 + V]$
corresponds to the Schr\"{o}dinger equation of Eq. (\ref{cla}).
The expectation value is
\beq \label{eq:mean_H}
	\langle H \rangle = 2c\sigma - \frac{\sigma b d}{2\sqrt{\pi k^{2}(2\sigma + \frac{1}{k^{2}})^3}} \ge E_{ground}.
\eeq
An extremizing variational value of $\sigma$ is given by
\beq \label{eq:sigma}
	\sigma = \frac{1}{2} \left(\left(\frac{3bd}{8\sqrt{\pi}ck^{4}}\right)^{2/5}-\frac{1}{k^{2}}\right).
\eeq
Substituting the above equation into Eq.~(\ref{eq:mean_H}), we obtain
\beq \label{eq:minE}
	\langle H \rangle = c(\beta-\frac{1}{k^{2}}) -  bd \left(\frac{\beta k^2-1}{4\sqrt{\pi}\beta^3 k^4}\right),
\eeq
where $\beta = \left(\frac{3bd}{8\sqrt{\pi}c k^{4}}\right)^{2/5}$. Similar to the earlier two cases of boundary deformations, the effective transition temperature can be found by approximating $a(T) \approx\alpha\, (T-T_c^{0})$.
The transition temperature
\beq \label{eq:TeffDislocation}
	T_c^{eff} =   T_c^{0} - \frac{1}{\alpha}c\left( \left(\beta- \frac{1}{k^{2}} \right) + bd \left(\frac{\beta k^{2} -1}{4\sqrt{\pi}\beta^3 k^{4}}\right)\right).
\eeq

\begin{figure}[hbt]
	\includegraphics[width=0.45\textwidth]{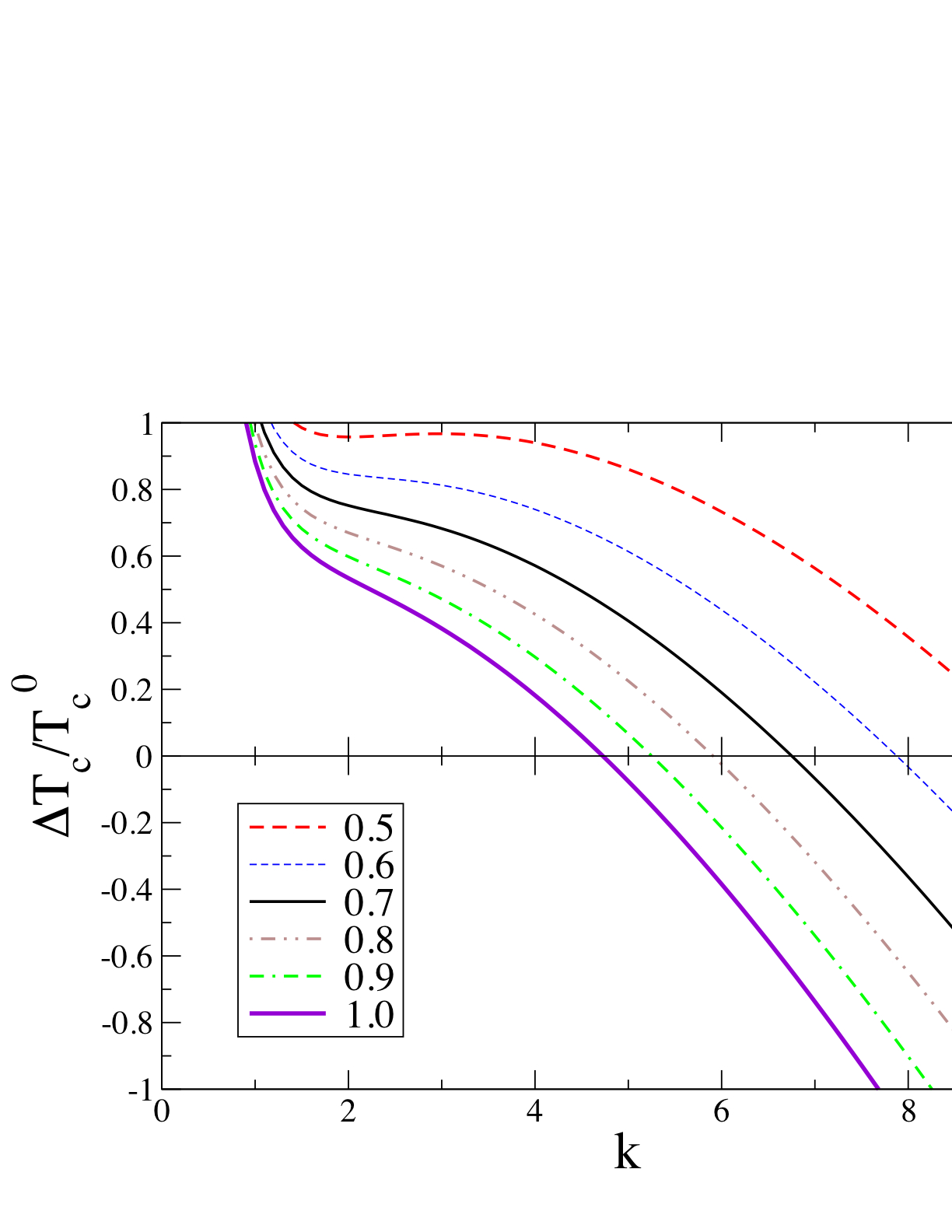}
	\caption{(Color online.) The shift in the local transition temperature in the vicinity of the dislocation core
	for different values of Burgers vector $b=0.5, \dots, 1.0$ as a function of core size $k$. The other parameters are set as $c=d=1$ and $\alpha
	= 1/T_{c}^{0}$.}
	\label{delta_dislocation}
\end{figure}

	In Fig.~\ref{delta_dislocation}, we plot $\Delta T_{c}=T_c^{eff}-T_c^0$ for different Burgers vectors $b$  as a
	function of  dislocation core size $k$ for fixed values of the parameters in the GL functional
	of Eqs.~(\ref{Fc}) and (\ref{lambda}): $c=d=1$.  Depending on the choice of parameters an increase or decrease in the local transition temperature relative to a bulk transition is possible. Assuming a typical Burgers vector magnitude of $b=1$ for an edge dislocation in an hcp crystal, with parameters $c=d=1$, a core size radius of $R=k/\sqrt{2} < 3.5$ lattice constants will result in a supersolid dislocation core prior to the bulk supersolid transition.
	
\section{Conclusions}
\label{conc}
In summary, we find that elastic deformations in supersolid lead to local changes in the transition temperature. For a positive coupling constant $d$ in Eq.~(\ref{lambda}) we obtain the results:
\begin{enumerate}
\item Edge contraction increases the supersolid transition temperature at and near the edges.
\item Edge expansion decreases the supersolid transition temperature at and near the edges.
\item The local supersolid transition may be enhanced or suppressed near a dislocation core. 
\end{enumerate}
This implies the observation of interesting effects. For example, for edge contractions, we would find that, below a certain temperature that is higher than the supersolid transition temperature, a sample of supersolid would have its supersolid edges partially decouple from its bulk crystal. 
Of course, the effects of elastic deformations on the supersolid transition are not limited to the few selected cases studied here. For example, the same physics applies to point defects like interstitials and vacancies, as well as to extended defects like grain boundaries and inclusions or voids.
The above conclusions were based
on the assumption of a positive coupling constant $d$ in Eq.~(\ref{lambda}).
Formally, for negative $d$, our conclusions would have been inverted-
an expansion would enhance the local supersolid transition
while a contraction would reduce the supersolid transition
temperature.  

Similar effects are found elsewhere in regions that locally expand or contract. 
In Ref.~\onlinecite{znmc} it was shown how 
a dislocation condensate may generally enhance and trigger    
superfluid behavior via a Higgs type mechanism. 
The presented GL approach of elastic deformations on the supersolid transition temperature is quite general and
also applies to superconductors. In fact,  dislocation defects and lattice-mismatched interfaces in superconductors
are known to create nonuniform strain and changes to the superconducting transition temperature, which have been studied extensively.\cite{Fogel2002}
Thus, our calculations of the changes in the local transition temperature due to a nonuniform elastic strain coupling in a Ginzburg-Landau approach are not limited to supersolidity and may as well apply to superconductivity.

\section{Acknowledgments}
This work was partially supported by the Center for Materials
Innovation (CMI) of Washington University, St.\ Louis and
by the US Dept.\ of Energy at Los Alamos National Laboratory
under contract No.~DEAC52-06NA25396.
We are grateful to A. T. Dorsey, J. Beamish, J. C. Davis, and J.-J. Su for many stimulating discussions.


\begin{thebibliography}{10}
\bibitem{quadrature}
Eq.~(\ref{EL}) could also be solved in the presence of the cubic term,
albeit not giving any new insight relative
to our linearized differential equation of Eq.~(\ref{cla}).
Throughout this work, we will assume a system that is symmetric about the origin, with a unidirectional axis of length $L$ along
which the displacements occur ($- L/2 \le z \le L/2$).
For real $\psi$, with boundary condition $\psi(z= - L/2) =0$,
multiplying Eq.~(\ref{EL}) by $\psi$ and integrating, 
we arrive at the general implicit relation
\begin{eqnarray}
z(\psi) =  - \frac{L}{2}  + \int_{0}^{\psi}  \frac{d \tilde{\psi}}{\tilde{\psi}} \sqrt{- \frac{c}{a + b \tilde{\psi}^{2} + \lambda(z)}},
\label{trivz}
\end{eqnarray}
for $z<0$.  Eq.~(\ref{trivz}) may be inverted to determine $\psi(z)$ for negative $z$.
For $z>0$, in a system with displacements that are symmetric
about  $z=0$, ~$\psi(z) = \psi(-z)$.
\end{thebibliography}
\end{document}